\documentclass[10pt,prd,twocolumn,showpacs,preprintnumbers]{revtex4}

\usepackage{graphicx,amsmath,amssymb,mathrsfs,dsfont,slashed}

\renewcommand{\Re}{\mathop{\mathrm{Re}}}
\renewcommand{\Im}{\mathop{\mathrm{Im}}}
\renewcommand{\b}[1]{\mathbf{#1}}
\renewcommand{\c}[1]{\mathcal{#1}}
\renewcommand{\u}{\uparrow}
\renewcommand{\d}{\downarrow}

\begin{document}

\def\nn{\nonumber}
\def\kc#1{\left(#1\right)}
\def\kd#1{\left[#1\right]}
\def\ke#1{\left\{#1\right\}}
\renewcommand{\Re}{\mathop{\mathrm{Re}}}
\renewcommand{\Im}{\mathop{\mathrm{Im}}}
\renewcommand{\b}[1]{\mathbf{#1}}
\renewcommand{\c}[1]{\mathcal{#1}}
\renewcommand{\u}{\uparrow}
\renewcommand{\d}{\downarrow}
\newcommand{\bsigma}{\boldsymbol{\sigma}}
\newcommand{\blambda}{\boldsymbol{\lambda}}
\newcommand{\Tr}{\mathop{\mathrm{Tr}}}
\newcommand{\sgn}{\mathop{\mathrm{sgn}}}
\newcommand{\sech}{\mathop{\mathrm{sech}}}
\newcommand{\diag}{\mathop{\mathrm{diag}}}
\newcommand{\Pf}{\mathop{\mathrm{Pf}}}
\newcommand{\half}{{\textstyle\frac{1}{2}}}
\newcommand{\sh}{{\textstyle{\frac{1}{2}}}}
\newcommand{\ish}{{\textstyle{\frac{i}{2}}}}
\newcommand{\thf}{{\textstyle{\frac{3}{2}}}}
\newcommand{\SUN}{SU(\mathcal{N})}
\newcommand{\N}{\mathcal{N}}
\newcommand{\be}{\begin{equation}}
\newcommand{\ee}{\end{equation}}
\newcommand{\tr}{\mathop{\mathrm{tr}}}
\newcommand{\sign}{\mathop{\mathrm{sign}}}

\title{Holographic fractional topological insulators in $2+1$ and $1+1$ dimensions}

\preprint{IPMU10-0157}

\author{Andreas Karch}
\affiliation{
Department of Physics, University of Washington, Seattle, WA
98195-1560, USA}

\author{Joseph Maciejko}
\affiliation{
Department of Physics, Stanford University, Stanford, CA 94305, USA\\
and\\
Stanford Institute for Materials and Energy Sciences,\\
SLAC National Accelerator Laboratory, Menlo Park, CA 94025, USA}

\author{Tadashi Takayanagi}
\affiliation{Institute for the Physics and Mathematics of the Universe (IPMU), University
of Tokyo, Kashiwa, Chiba 277-8582, Japan}

\date\today

\begin{abstract}
We give field theory descriptions of the time-reversal invariant quantum spin Hall insulator in $2+1$ dimensions and the particle-hole symmetric insulator in $1+1$ dimensions in terms of massive Dirac fermions. Integrating out the massive fermions we obtain a low-energy description in terms of a topological field theory, which is entirely determined by anomaly considerations. This description allows us to easily construct low-energy effective actions for the corresponding `fractional' topological insulators, potentially corresponding to new states of matter. We give a holographic realization of these fractional states in terms of a probe brane system, verifying that the expected topologically protected transport properties are robust even at strong coupling.
\end{abstract}


\maketitle

\section{Introduction}

Recent years have witnessed tremendous activity in the application of the anti-de Sitter (AdS)/conformal field theory (CFT) correspondence~\cite{Maldacena:1997re,Gubser:1998bc,Witten:1998qj} to the study of condensed matter systems~\cite{AdSCMT}. Conformal symmetry (or the less restrictive Galilean~\cite{son2008,balasubramanian2008} or Lifshitz~\cite{kachru2008} symmetries) on the field theory side of the correspondence implies that one is describing a critical point or a gapless phase of matter. In the latter case, if some of the gapless degrees of freedom are charged under electromagnetism, one is describing a metal or a superconductor, while if all electrically charged degrees of freedom are gapped and absent from the low-energy spectrum, one is describing an insulator or a superfluid. In the case of an insulator, the electrically neutral gapless degrees of freedom could correspond, for example, to low-energy phonons.

A novel class of insulators, time-reversal ($T$) invariant topological insulators~\cite{TIs} (TI), has been recently theoretically predicted~\cite{Kane2005,bernevig2006a,bernevig2006d,Fu2007,Zhang2009} and subsequently observed~\cite{konig2007b,roth2009,Hsieh2008,Xia2009,Chen2009} experimentally. The spectrum of electrically charged excitations in a TI is gapped in the bulk, as one expects for an insulator, but is gapless on the boundary. $T$-invariance ensures the crossing of the energy-momentum dispersion relation of the boundary states at certain momenta, such that the spectrum of a topologically nontrivial insulator with gapless boundary states cannot be adiabatically deformed to that of a topologically trivial insulator without gapless boundary states~\cite{wu2006,xu2006}. The perturbative stability of the boundary states against arbitrary $T$-invariant perturbations is protected by a bulk topological invariant which, for $T$-invariant TI in $3+1$ and $2+1$ dimensions, is $\mathbb{Z}_2$-valued~\cite{Kane2005,roy2009,Fu2007,roy2009b,moore2007,Qi2008}. The consideration of other discrete symmetries such as charge conjugation/particle-hole symmetry ($C$) has led to the theoretical prediction of many more classes of topological insulators~\cite{schnyder2008,Kitaev2009,schnyder2009,Ryu2009}, among them the $C$-invariant topological insulator in $1+1$ dimensions which admits a $\mathbb{Z}_2$ classification.

The topological classification mentioned above relies for the most part on the band theory of noninteracting fermions. Although it is expected to hold for weak enough interactions, strong interactions may qualitatively change the picture. For example, it was shown in Ref.~\onlinecite{fidkowski2010,fidkowski2010b,turner2010} that the $\mathbb{Z}$ classification of the BDI symmetry class in $1+1$ dimensions breaks down to $\mathbb{Z}_8$ in the presence of interactions. A strongly interacting `fractional' generalization of the $T$-invariant TI in $2+1$ dimensions, also known as the quantum spin Hall (QSH) insulator, was considered recently~\cite{bernevig2006a,Levin2009b}. Finally, recent work~\cite{Maciejko:2010tx,Swingle:2010rf} identified the topological field theory for a fractional version of the $T$-invariant TI in $3+1$ dimensions as the topological $\theta$-term of an effective gauge theory.

The fractional TI phases mentioned in the previous paragraph are strongly interacting, rendering any explicit realization --- say, at the level of a microscopic Hamiltonian --- difficult. Assuming that the underlying theory leads to a fractionalization of electrons into partonic degrees of freedom, Ref.~\onlinecite{Maciejko:2010tx,Swingle:2010rf} constructed an effective theory of the partons together with gauge fields which ensure that the only gauge-invariant states carry the quantum numbers of the original electrons.  Even when these partonic degrees of freedom themselves are strongly correlated, the topological properties can still be uniquely determined as they are governed by an anomaly calculation. One may still worry that the partonic picture is inadequate when it itself is strongly coupled; it also would be desirable to embed this low-energy field theory in a self-consistent microscopic theory, not necessarily a lattice model. Fortunately, certain strongly coupled field theories can be given an explicit microscopic realization in terms of string theory, through the AdS/CFT correspondence. Very recently, such a holographic realization of the fractional $T$-invariant TI in $3+1$ dimensions was given in terms of Type IIB superstring theory on AdS$_5\times S^5$ with probe D7-branes~\cite{toappear}. In this work, we study the holographic realization of the $T$-invariant QSH insulator in $2+1$ dimensions and the $C$-invariant insulator in $1+1$ dimensions.

As our basic strategy for constructing a continuum field theory of the QSH effect is, in some respects, similar to the effective field theory description of the quantum Hall effect, let us briefly compare and contrast the two cases. In both theories, topological Chern-Simons terms are generated by integrating out massive fermions. While in the integer quantum Hall effect time-reversal invariance is broken, in our case we choose a particular combination of masses for pairs of fermions that preserves time-reversal invariance. In both cases, charge fractionalization can give rise to fractional versions of the effect. In the case of the quantum Hall effect, this is only one of many possible descriptions of the phenomenon, but it has been experimentally verified that these fractional phases do indeed exist in real materials. In our case, it is not clear whether the charge fractionalization can be realized in an actual material. We are however able to demonstrate two important points. Firstly, consistent low-energy effective theories that give rise to quantized fractional QSH transport can be constructed. Therefore, there is no fundamental principle that forbids such fractional states, even though at the moment we have no clear answer to the question whether they can be realized via a system of interacting electrons. Secondly, we are able to demonstrate the existence of UV complete theories that do give rise to the low-energy physics we study: systems of intersecting branes in string theory.

This paper is organized as follows. In Sec.~\ref{sec:FT_QSHI}, we review the field theory description of $T$-invariant TI in $3+1$ and $2+1$ dimensions in terms of massive Dirac fermions~\cite{Qi2008} and the Adler-Bell-Jackiw anomaly, for both the `noninteracting' ($\mathbb{Z}_2$) and fractional cases. In Sec.~\ref{sec:H_QSHI}, we give a holographic realization of the fractional QSH insulator in terms of probe D5-branes in Type IIB superstring theory on AdS$_5\times S^5$. In Sec.~\ref{sec:FT_PHSI}, we describe the class D and class DIII $C$-invariant TI in $1+1$ dimensions, once again in terms of massive Dirac fermions and the chiral anomaly. In Sec.~\ref{sec:H_PHSI}, we give a holographic realization of a fractional $C$-invariant TI in $1+1$ dimensions in terms of 6ND (for class D) and 4ND (for class DIII) probe brane setups. In both $2+1$ and $1+1$ dimensions, the topological response properties of the corresponding fractional TI states follow directly from the topological Wess-Zumino (WZ) terms in the D-brane actions.

\section{Field theory of the quantum spin Hall insulator}\label{sec:FT_QSHI}
\subsection{Topological insulator in $3+1$ dimensions}

We first recall some basic facts about the field theory description of the $T$-invariant TI in $3+1$ dimensions~\cite{Qi2008}. We use the conventions of Ref.~\onlinecite{Peskin:1995ev}. In particular, this means that we are working in mostly minus metric signature. The action for a single massive Dirac fermion $\psi$ is
\be
\label{freeaction}
{\cal L} = \overline{\psi} (i\gamma^{\mu} \partial_{\mu} - M) \psi,
\ee
where $M$ is the Dirac mass, $\overline{\psi} \equiv \psi^{\dagger} \gamma^0$ is the Dirac conjugate, and $\gamma^\mu$, $\mu=0,\ldots,3$ are the Dirac matrices. The mass parameter $M$ in general can be taken to be complex if we treat $\overline{\psi} \psi$ and $\overline{\psi} \gamma^5 \psi$
as the real and imaginary parts of the fermion bilinear, with $\gamma^5\equiv i\gamma^0\gamma^1\gamma^2\gamma^3$. As the former is even under $T$, whereas the latter is odd under $T$, $T$ is only a good symmetry if $M$ is real. The mass term explicitly breaks the chiral symmetry associated with individual rotations of $\psi_L = P_L \psi$ and $\psi_R = P_R \psi$, where we defined the projectors $P_{L/R} \equiv \frac{1}{2} ( 1 \pm \gamma^5)$.
The broken chiral symmetry can be used to rotate the phase of $M$ to zero, so that the mass can always be chosen to be real and positive. Due to the chiral or Adler-Bell-Jackiw anomaly, however, one will generate a $\theta$-angle equal to the phase of the original mass term. For a $T$-invariant theory the original mass had to be real, but could be positive or negative. Therefore, the generated $\theta$ has to be $0$ or $\pi$.

It was shown in Ref.~\onlinecite{Maciejko:2010tx} that this discussion can easily be generalized to fractional TI. If we demand that the electron breaks up into $N$ partons of charge $1/N$ (in units of the electron charge $e$) each one with a real, negative mass, the chiral anomaly yields a $\theta$-term with angle $\theta=\mathcal{C} \pi$ when we rotate all masses to be real and positive, where $\mathcal{C}$ is the sum over the electric charge squared of all fields. One therefore obtains $\mathcal{C} = N \cdot (1/N)^2 = 1/N$ for $N$ partons. In order to ensure that outside the TI the only physical states are the electrons, we need to add a `statistical' gauge field which ensures that the only gauge-invariant states carrying electric charge are the electrons composed of $N$ partons. In vacuum the statistical gauge field needs to be in a confined phase, so that the partons are always tightly bound into electrons. Inside the TI, the partons are allowed to deconfine. Ref.~\onlinecite{Maciejko:2010tx,Swingle:2010rf} described both Abelian and non-Abelian models realizing this scenario. The former naturally appear in deconfined phases. For the latter, one needs to add extra light matter to drive the $SU(N)$ gauge field into a deconfined phase. A nice example is to chose ${\cal N}=4$ supersymmetric Yang-Mills (SYM) theory as the statistical gauge sector inside the insulator. This theory is conformal and so indeed in a deconfined phase~\cite{seiberg1988}. The extra massless adjoint matter is electrically neutral, so we do describe an insulator, albeit in the presence of extra light `phonons'. While details of the model, such as the ground state degeneracy, can depend on the exact properties of the statistical gauge field, the quantized value of $\theta$ does not. It is uniquely determined by the anomaly.

An interface across which the mass of a Dirac fermion crosses from real and positive to real and negative localizes a single massless ($2+1$)-dimensional fermion~\cite{Jackiw:1975fn,callan1985}. No mass term can be added consistent with $T$-invariance~\cite{Jackiw:1980kv}. In $2+1$ dimensions the action of $T$ on the Dirac field is
$T \psi T^{-1} = \sigma_2 \psi$ and hence $T \overline{\psi} T^{-1} = - \overline{\psi}\sigma_2$. Therefore, the standard $\overline{\psi} \psi$ mass term is $T$-odd. One can however add a mass term for pairs of fermions consistent with $T$-invariance. In fact, dimensional reduction of the ($3+1$)-dimensional $T$-invariant mass term for a single Dirac fermion $\Psi=(\psi_1,\psi_2)$ becomes a $T$-invariant mass for two fermions in $2+1$ dimensions~\cite{Jackiw:1980kv}:
\be
\label{threedmass}
{\cal L}_M = -M \overline{\Psi} \Psi = -M (\overline{\psi}_2 \psi_2 - \overline{\psi}_1 \psi_1),
\ee
where $M$ is a real parameter.
Therefore, the ($2+1$)-dimensional surface modes of the ($3+1$)-dimensional $T$-invariant TI can be removed in pairs~\cite{Fu2007,Qi2008}. If we start with $N_f$ equal-mass Dirac fermions in $3+1$ dimensions and consider an interface across which their common mass changes sign, we localize $N_f$ massless ($2+1$)-dimensional fermions on the interface. They can however pair up and become massive via perturbations of the form displayed in Eq.~(\ref{threedmass}). Therefore, only if $N_f$ is odd is one guaranteed to have a single massless fermion living on the interface even in the presence of ($T$-invariant) perturbations. This is the only sense in which the continuum description knows about the difference between a $\mathbb{Z}$ classification versus a $\mathbb{Z}_2$ classification. The free fermions are classified by an integer $N_f$. However, including disorder we should allow for the most general mass perturbation to be added consistent with $T$-invariance. As a generic mass term can remove surface modes in pairs, $N_f$ is not a topological quantum number. But the $\mathbb{Z}_2$ distinction whether $N_f$ is even or odd is topological.

The Lagrangian describing the localized fermions has a $U(1) \times SU(N_f)$ global symmetry. Note that the $T$-allowed mass term is neutral under the diagonal $U(1)$, but transforms non-trivially under the non-Abelian $SU(N_f)$ flavor group. This symmetry property will allow us later to identify the corresponding perturbation in the brane realization of (fractional) TI and the QSH insulator.

\subsection{Quantum spin Hall insulator}

To describe in the same spirit a (fractional) TI in $2+1$ dimensions, we need to start with ($2+1$)-dimensional fermions with a $T$-invariant mass. As just discussed, this requires an even number of fermions. The simplest example is to take one flavor of a pair of fermions; for the general case we should allow $N_f$ pairs. The massless theory has a global $U(1) \times SU(2 N_f)$ symmetry, as all fermions are equivalent and can be rotated into one another. The $T$-invariant mass term from Eq.~(\ref{threedmass}) breaks this global symmetry to $U(N_f) \times U(N_f)$, as half of the fermions have positive mass while the other half have negative mass. The diagonal $U(1)$ subgroup is electromagnetism and all fermions carry charge $+1$ under it. The second $U(1)$, under which half the fermions carry charge $+1$ and the other half charge $-1$, is a global symmetry. If this fermion system is part of a supersymmetric gauge theory, this global $U(1)$ is typically an $R$-symmetry; this will in particular be true in the brane embeddings we study below. Hence we will from now on refer to it as $U(1)_R$. In the condensed matter context however, this $U(1)$ plays a very different role; it is the ($z$ component of the) spin of the particle~\cite{Rsymmetry}. For free, non-relativistic fermions, spin is simply a global symmetry not tied to the space-time properties of the fermions. In particular, the $z$-component of the spin is a global $U(1)$ symmetry under which all the fermions carry charge. These are exactly the properties of the $U(1)_R$ symmetry we introduced above. Therefore, $R$-charge transport is the continuum quantity that corresponds to what is referred to as `spin' transport in the condensed matter literature. To realize the QSH effect in the relativistic continuum theory, we need to find a realization of a `quantum $R$-Hall effect', i.e. an $R$-current induced in the $y$ direction in response to an electric field pointing in the $x$ direction.

Note that the $T$-invariant mass term in the Lagrangian is once more characterized by a real parameter $M$. Just as in $3+1$ dimensions, whether the mass $M$ is positive or negative corresponds to topologically distinct situations. That is, we can not smoothly deform one into the other without closing the gap. Note that the two fermions have masses of opposite sign, so there is always one fermion with positive mass and one with negative mass. We are therefore free to define the topologically trivial insulator (including vacuum) as the situation in which the positive $R$-charge fermion has positive mass, whereas a negative mass for the positive $R$-charge fermion corresponds to a topologically non-trivial insulator. We can integrate out the massive fermions; their topological properties are then simply encoded in the Chern-Simons terms that are generated~\cite{Niemi:1983rq,Redlich:1983kn,Redlich:1983dv}. As we integrate out a fermion with flavor label $i$ with mass $M_i$ carrying charge $q_{i,a}$ under a global or local $U(1)$ symmetry labeled by $a$, we generate a (mixed) Chern-Simons term of level
\be\label{CSlevel}
k_{ab} = \frac{1}{2} \sum_i q_{i,a} q_{i,b} \, \sgn(M_i).
\ee
From Eq.~(\ref{CSlevel}) it is clear the contributions to the $A \wedge dA$ Chern-Simons term cancel between the two fermions: while they have the same charge squared of $1$, they have opposite sign mass. The same holds for the $A^R \wedge dA^R$ Chern-Simons term for the non-dynamical background vector potential $A^R$ that one may want to introduce to describe non-trivial non-dynamical $U(1)_R$ background fields. We do however generate a non-trivial mixed Chern-Simons term $A^R \wedge dA$~\cite{Rsymmetry2}. As the two fermions have charges of same sign under the Maxwell $U(1)$ but charges of opposite sign under the $U(1)_R$ in addition to their opposite sign of $M$, this time the two contributions add and we generate a Chern-Simons term at level $1$. This is exactly the QSH effect. In a background electric field, the $R$-current we obtain by varying the action with respect to the $R$-charge gauge field carries one Hall quantum of $R$-current in the $y$ direction from the unit level Chern-Simons term. For $N_f$ fermion pairs we get a Chern-Simons term at level $N_f$. However, once again the only topologically protected information is whether $N_f$ is even or odd; an even number of pairs can be removed in a $T$-invariant way without generating any Chern-Simons terms.

On a boundary between a topologically non-trivial ($2+1$)-dimensional insulator and a topologically trivial insulator such as vacuum, one again localizes massless fermions, this time a helical edge state. This helical state consists of two ($1+1$)-dimensional chiral fermions of opposite chirality and opposite $R$-charge. Unlike the chiral edge modes of the quantum Hall effect, here a mass term can be written down corresponding to backscattering of left- and right-movers. However, as in the ($3+1$)-dimensional case with the ($2+1$)-dimensional massless surface mode, the mass term is $T$-odd and so will not be generated by arbitrary perturbations as long as they preserve $T$-invariance. A quartic backscattering term is allowed but is irrelevant at the free fermion fixed point~\cite{wu2006,xu2006,maciejko2009}. The edge modes do not contribute to charge transport, but do transport $R$-charge (i.e. the $z$ component of spin). For $N_f$ topologically non-trivial fermion pairs in $2+1$ dimensions, one correspondingly finds $N_f$ helical edge-mode pairs. Again, an even number of them can be made massive by adding a $T$-invariant mass term, hence we only have a $\mathbb{Z}_2$ TI.

While most of this is well known, the identification of spin-transport with an $R$-current allows us to realize this phenomenon in some of the best understood strongly coupled ($2+1$)-dimensional field theories: supersymmetric gauge theories and their brane realizations.

\subsection{Fractional quantum spin Hall insulator}

Just as in the example of the ($3+1$)-dimensional TI of Ref.~\onlinecite{Maciejko:2010tx}, it is straightforward to generalize this continuum picture of a QSH insulator to a continuum description of a fractional QSH system. Again, we simply demand that the electron is allowed to fractionalize into $N$ partons. Additional gauge fields should be added to ensure that the electrons are the only gauge-invariant states, just as in $3+1$ dimensions. The coefficient of the Chern-Simons term encoding the spin or $R$-transport is once more completely insensitive to the details of the gauge sector, and given by ($i$ running over all fields)
\be
k = \frac{1}{2} \sum_{i} q^{\text{electric}}_{i} q^R_{i} \, \sgn(M_i) =\frac{1}{N},\nonumber
\ee
where we have used the fact that in the non-trivial insulator all partons have negative mass $M$ and carry both spin (i.e. $R$-charge) and Maxwell charge of magnitude $1/N$, with same-sign $R$-charge but opposite-sign Maxwell charge for the two fermions in a $T$-invariant pair.

\section{Holographic fractional quantum spin Hall insulator}\label{sec:H_QSHI}

\subsection{Bulk theory}

\begin{figure}[t]
\begin{center}
\includegraphics[scale=0.5,angle=0]{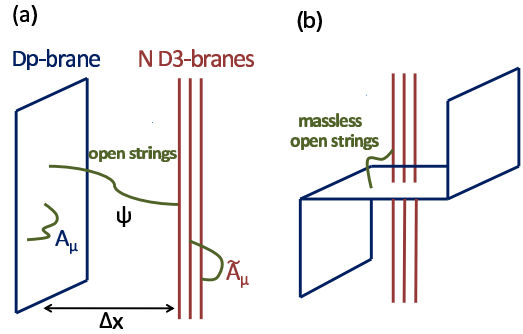}
\end{center}
\caption{Geometry of the basic construction underlying holographic realizations of the various topological insulators we describe. (a) General D3-D$p$ system, where the $N$ D3-branes realize the $\mathcal{N}=4$ super Yang-Mills sector containing the statistical $SU(N)$ gauge field $\tilde{A}_{\mu}$, and matter Dirac fermions $\psi$ are added via inclusion of $p=5$ and $p=3$ ``flavor" D$p$-branes containing the non-dynamical $U(1)$ gauge field $A_\mu$. A finite D3-D$p$ separation $\Delta x\neq 0$ corresponds to a finite mass $|M|=T_s\Delta x$ for the $\psi$ fermions, where $T_s$ is the string tension. (b) Configuration with a mass term that interpolates between two topologically distinct phases. As the D3-branes and the D$p$-brane now intersect, massless fermionic matter is localized at the intersection.}
\label{brane}
\end{figure}

In order to realize the continuum field theory of the QSH effect in terms of a brane system in string theory and give a holographic description, it is easiest to first embed this sector in a supersymmetric gauge theory. As in the holographic realization of the ($3+1$)-dimensional TI, a good statistical gauge sector is ${\cal N}=4$ SYM theory with $SU(N)$ gauge group. This theory naturally describes a deconfined phase of the non-Abelian gauge field. At strong coupling it has a very simple holographic dual. As in the ($3+1$)-dimensional case~\cite{toappear}, we add the partons as a $\mathcal{N}=2$ supersymmetric hypermultiplet preserving 8 supersymmetries, containing a single massive pair of ($2+1$)-dimensional fermions as well as scalar superpartners that do not contribute to the anomaly. We constrain these charge carriers to live on a ($2+1$)-dimensional defect, while they interact with the `phonon' bath made of ${\cal N}=4$ SYM fields propagating in the full $3+1$ dimensions. The brane picture underlying all the holographic constructions in this paper is depicted in panel (a) of Fig.~\ref{brane}. This basic setup is similar to Ref.~\onlinecite{Ryu:2010hc,Ryu:2010fe}, where weakly interacting topological insulators have been given in terms of D$p$-D$q$ systems. However, in our construction we consider multiple D3-branes in order to construct holographic duals by taking the large $N$ limit and to
realize fractional TI. From the discussion above, this system will give rise to a QSH current governed by a mixed $R$/Maxwell Chern-Simons term of level $1/N$. In the holographic calculation it is much more convenient, for the purpose of counting powers of $N$ in the large-$N$ limit that underlies the holographic calculation, to assign charge $1$ to the parton, giving the electron a total charge $N$. In this case the anomaly argument predicts a mixed Chern-Simons term of level
\be
\label{fieldtheoryprediction}
k = N.
\ee
This change in normalization was already needed in the holographic realization of the ($3+1$)-dimensional TI. However, this still corresponds to the same fractional QSH state and is merely a matter of convention.

In string theory, the $N$ ($2+1$)-dimensional defect partons in the fundamental ($N$) representation of $SU(N)$ can be added to the ${\cal N}=4$ SYM phonon bath by adding $N_f$ D5-branes to a system of $N$ D3-branes, where the latter realize the ${\cal N}=4$ SYM system. In the weak coupling limit, where the same system is best thought of as the field theory living on the worldvolume of intersecting branes, the D3- and D5-branes occupy directions as indicated in Table~\ref{tableone}.
\begin{table}
\begin{tabular*}{0.23\textwidth}{l|l}
D3&$\times$\hspace{1mm}$\times$\hspace{1mm}$\times$\hspace{1mm}$\times$
\put(5.9,2.3){\circle{5}}\put(15.9,2.3){\circle{5}}\put(25.9,2.3){\circle{5}}\put(35.9,2.3){\circle{5}}\put(45.9,2.3){\circle{5}}\put(55.9,2.3){\circle{5}}\\
\hline
D5&$\times$\hspace{1mm}$\times$\hspace{1mm}$\times$\hspace{1mm}
\put(5.9,2.3){\circle{5}}
\hspace{12.5pt}$\times$\hspace{1mm}$\times$\hspace{1mm}$\times$
\put(6,2.3){\circle{5}}\put(16,2.3){\circle{5}}\put(26.2,2.3){\circle{5}}
\end{tabular*}
\caption{The D3-D5 system. Crosses indicate directions occupied by the respective brane, circles indicate transverse directions.}
\label{tableone}
\end{table}
The symmetries of the system are manifest geometrically. In addition to Poincar\'{e} invariance along the $2+1$ dimensions represented by $012$, we have an $SO(3) \times SO(3)$ global $R$-symmetry associated with rotations in the $456$ and $789$ directions, respectively. Giving the hypermultiplets a finite mass corresponds to moving the D5-branes away
from the stack of D3-branes in the $x_9$ direction. Correspondingly, the second $SO(3)$ factor is broken to $SO(2) \cong U(1)_R$. It is this last $U(1)_R$ factor that plays the role of the $z$ component of spin in our system.

At strong coupling, the theory on the $N$ D3-branes is dual to Type IIB supergravity on AdS$_5 \times S^5$. Incorporation of the defect fermions proceeds via embedding a D5-brane~\cite{Karch:2000gx} wrapping an $S^2$ inside the $S^5$ and extending along an AdS$_4$ slice of AdS$_5$. The dynamics of the brane is governed by minimization of its worldvolume. In addition, it has a WZ term in its action, coupling the worldvolume gauge fields to the form fields present in the background, in particular to the $N$ units of D3-brane flux that support the background geometry. These terms encode the anomalies and so should be solely responsible for the QSH current, as we will confirm below. We write the metric on the sphere $S^5$ as
\be
ds^2_{S^5} = \cos^2 \tilde{\theta} \,  d \Omega_2^2 + d \tilde{\theta}^2 + \sin^2 \tilde{\theta} \, d \tilde{\Omega}_2^2, \label{sphereone}
\ee
where the metric on the $S^2$ wrapped by the brane is denoted $d \Omega_2^2$.
The brane embedding is given by a function $\tilde{\theta}(r)$, where $r$ is the radial coordinate in
AdS$_5$ when the metric is written in the standard form
\be
ds^2_{\mathrm{AdS}_5} =  \frac{dr^2}{r^2} + r^2 (-dt^2 + dx^2 + dy^2 + dz^2).
\nonumber
\ee
For a mass $M$ hypermultiplet, $\tilde{\theta}=\arcsin(|M|/r)$ is
an exact solution. It describes a brane that approaches $\theta=0$ at large $r$ (close to the boundary of
AdS space), but terminates smoothly at $r=|M|$. The brane sits at a fixed position on the second sphere, the metric of which is denoted by $d \tilde{\Omega}_2^2$.
Positive and negative $M$ embeddings correspond to a D5-brane sitting at opposite poles of this sphere. $SO(2)\cong U(1)_R$ rotations leave the position of the brane on this sphere invariant, and so are symmetries of the system. The QSH current calculated above using simple anomaly arguments should be reproduced via the WZ terms in the brane action.

The $R$-symmetry gauge field $A^R_{\mu}$ appears in the WZ terms as a component of the Ramond-Ramond (RR) 4-form field $C_4$, as worked out in detail in Ref.~\onlinecite{Aharony:1999rz}.
The WZ term containing $C_4$ reads
\begin{eqnarray}
S_\text{WZ}&=& (2 \pi \alpha') T_5 \int  F \wedge C_4 , \quad T_5 = \frac{1}{ (2 \pi)^5 (\alpha')^3} .
\label{conventions}
\end{eqnarray}
In order to directly apply the formulas obtained in Ref.~\onlinecite{Aharony:1999rz}, one needs to use a slightly
different parametrization of $S^5$,
\be
\label{ofermetric}
ds^2_{S^5} = \cos^2 \theta \,  (d \chi^2 + \cos^2 \chi \, d \Omega_2^2) + d\theta^2 + \sin^2 \theta \, d \psi^2.
\ee
It is easy to check that with an ansatz $\theta=\theta(r)$, $\chi=\chi(r)$ we get a solution to the minimal-area equation of motion which reads $\theta=0$, $\chi=\arcsin(|M|/r)$. This parametrization of the sphere makes the
$U(1)_R$ symmetry manifest as the shift symmetry in $\psi$. According to Ref.~\onlinecite{Aharony:1999rz} one has
\be
\label{oferrfield}
C_{4,\mu abc} = \tilde{\eta} A^R_{\mu} \omega_{abc}, \ee
with~\cite{normalization}
\be
\label{oferconvention}
\tilde{\eta} = 8 \pi \alpha'^2 N,\ee
where the index $\mu$ runs over $t$, $x$ and $y$ and $\omega_{abc}$ stands for the volume form of the 3-sphere in $S^5$.
In order to evaluate the contributions of these terms, it is useful to change the parametrization of the brane.
Instead of using $r$ as a worldvolume coordinate and describing the embedding by a function $\theta(r)$, we can directly use $\theta$ as one of the worldvolume coordinates. For massive embeddings the range of the
$\theta$ integral goes from $0$ (out at the boundary) to $\pi/2$ (where the brane ends). In the parametrization of Eq.~(\ref{ofermetric}), $\theta$ covers the range from $-\pi/2$ to $\pi/2$, so we see that for a massive embedding $\theta$ sweeps out {\it half} of the 3-sphere. Performing the integral over this 3-hemisphere of area $\pi^2$,
the WZ term reduces to a Chern-Simons term on the boundary:
\begin{eqnarray}
\nonumber
S_\text{WZ} &=&  \tilde{\eta} T_5 (2 \pi \alpha') \pi^2 \int dt\,dx\,dy\,A^R \wedge F \\
 &=&
\frac{N}{4 \pi} \int dt\,dx\,dy\left ( A^R \wedge F + A \wedge F^R \right ).\label{CSterm}
\end{eqnarray}
As the prefactor of the Chern-Simons term should be of the form of an integer level divided by $4 \pi$, we see that the supergravity calculation is in perfect agreement with our field theory prediction Eq.~(\ref{fieldtheoryprediction}). As expected, the Chern-Simons terms are independent of details of the embedding function. The WZ terms correctly encode the anomaly in the field theory.

\subsection{Surface theory and mass deformations}

In order to holographically realize the interface between a topologically trivial and a non-trivial insulator, one needs to construct the brane embedding dual to a hypermultiplet not with constant mass but with a mass $M(x)$ that depends on one of the spatial directions and interpolates between a positive and a negative value of $M$ at $x=\pm \infty$, respectively, as depicted in panel (b) of Fig.~\ref{brane}.  Such an embedding has been constructed numerically and, for small $x$, in a series expansion for the D3-D7 system~\cite{toappear}. The same methods can also be applied to the D3-D5 system. Any such domain wall configuration will localize a helical edge state on the interface, as we have argued above.

Instead of repeating the details of this construction here, let us focus on a different aspect of the theory. As we reviewed above, for $N_f$ flavors of fermions with a topologically non-trivial mass term we will get $N_f$ localized helical edge states, just as one gets $N_f$ Dirac cones on the surface of a ($3+1$)-dimensional TI with $N_f$ massive Dirac fermions. This is obvious also from the probe brane calculation. For $N_f$ coincident probe branes the action describing the embedding is simply $N_f$ times that of a single probe brane, so we automatically get a Chern-Simons level of $N N_f$. At this level of description, the TI and the QSH effect seem to be classified by an integer $N_f$, instead of just a $\mathbb{Z}_2$  invariant (the sign of the mass). The important point here is that the appearance of $N_f$ helical edge modes is accidental, and mass terms can be added that remove these edge modes in pairs. The setup with $N_f$ coincident flavor branes is fine-tuned to have all $N_f$ edge modes massless, but after adding the most general $T$-invariant perturbation we are always left with either zero or one helical edge mode. The topologically robust question is only whether $N_f$ is even or odd.

What remains to be done is to identify the corresponding deformation in the holographic theory. For simplicity let us do this in the case of the ($2+1$)-dimensional surface modes of a ($3+1$)-dimensional TI. As reviewed above, in this case the theory of the $N_f$ surface modes has a $U(1) \times SU(N_f)$ global symmetry. The mass terms transform in the adjoint representation of this global symmetry; the $T$-invariant masses are the antisymmetric part of this. In the dual holographic theory we identified the field dual to the mass term with the position in the $\tilde{\theta}$ direction of the brane (or, in terms of the embedding space coordinates, the $x_9$ position of the brane). For $N_f$ coincident branes this parameter indeed gets promoted to a $N_f\times N_f$ $X_9$ matrix of fields that transforms in the adjoint representation of $U(N_f)$. The diagonal $U(1)$ part of this has a geometric interpretation in terms of the center-of-mass position of the brane. However, the $T$-invariant mass term corresponds to, as we argued above, turning on some of the non-trivial $SU(N_f)$ components of the field. Therefore we reassuringly find that the brane allows us to realize the $T$-invariant mass terms that can lift the degeneracy of all but one of the surface modes. In the D-brane classification of TI put forward in Ref.~\onlinecite{Ryu:2010hc,Ryu:2010fe}, the authors went one step further and imposed in addition a $\mathbb{Z}_2$  orientifold action on the brane setup that projected out all the $T$-non-invariant fields from the theory, which leads to the expected $\mathbb{Z}_2$ D-brane charge. Here we content ourselves with simply not allowing the corresponding perturbation to be turned on in the field theory Lagrangian.

\subsection{Plateau-like transition}

If we assume that the D5-brane extends in the direction $\tilde{\theta}=0$
near the AdS boundary and takes the maximum value $\theta_*$ $(0\leq \theta_*\leq \frac{\pi}{2})$,  we can read
off the relevant Chern-Simons coupling
\be
\frac{N}{2\pi}\sin\tilde{\theta}_*\int F\wedge A^R.
\nonumber
\ee
Thus we obtain the value of the spin Hall conductivity $\sigma^\text{spin}_{xy}$, which is defined by the ratio
of the spin current to the transverse electric field $\sigma^\text{spin}_{xy}=\frac{j^\text{spin}_{x}}{E_y}$:
\be
\sigma^\text{spin}_{xy}=\frac{N\sin\tilde{\theta}_*}{2\pi}.
\nonumber
\ee
At zero temperature, we find $\sin\tilde{\theta}_*=\sgn M$ as already mentioned, and thus
the value of $\sigma^\text{spin}_{xy}$ jumps at $M=0$.
As we have seen, in our model there are two phases corresponding to the positive and negative mass,
and they are distinguished by the sign of the spin Hall conductivity.
This is analogous to the plateau transition in the quantum
Hall effect. We refer the reader to Ref.~\onlinecite{Davis:2008nv,Fujita:2009kw,Bergman:2010gm} for holographic D-brane constructions of
(fractional) quantum Hall effects and the plateau transition.

Let us move on to finite temperature. For this we consider a probe D5-brane in the
Type IIB supergravity background given by the product of an AdS$_5$ black hole and $S^5$. We can write the metric of the AdS$_5$ black hole as
\be
ds^2=\frac{dr^2}{r^2h(r)}+r^2(-h(r)dt^2+dx^2+dy^2+dz^2),
\nonumber
\ee
where $h(r)=1-\frac{r_+^4}{r^4}$. The temperature of this black hole is given by $T=\frac{r_+}{\pi}$.
The Dirac-Born-Infeld Lagrangian of a probe D5-brane in the $S^5$ coordinate system of Eq.~(\ref{sphereone}) is given by
\be
L=r^2\cos^2\tilde{\theta}\sqrt{1+r^2h(r)\tilde{\theta}'^2}.\label{actionfin}
\ee
From a solution to the equation of motion obtained from Eq.~(\ref{actionfin}), we can read off
the mass from the behavior $\tilde{\theta}(r)\simeq \frac{M}{r}$ in the limit $r\to\infty$ and $\tilde{\theta}_*$ from the value of $\tilde{\theta}(r)$ at the horizon $r=r_+$. Notice that
since here we consider a black hole embedding, D5-brane solutions will end at the horizon.

To see the behavior near the critical point $M=0$, we can assume that $\tilde{\theta}$ is infinitesimally small. This approximation leads to the analytical solution
\be
\tilde{\theta}(r)=\tilde{\theta}_*\cdot {}_2F_1\left(\frac{1}{4},\frac{1}{2};1;1-\frac{r_+^4}{r^4}\right),
\nonumber
\ee
where $_2F_1(a,b;c;z)$ is the Gauss (ordinary) hypergeometric function. Therefore, the slope of the spin Hall conductivity $\sigma^\text{spin}_{xy}$ as a function of the mass behaves near the
critical point as
\be
\frac{\partial \sigma^\text{spin}_{xy}}{\partial M}=\frac{N}{2\pi}
\frac{\partial \sin\tilde{\theta}_*}{\partial M}\simeq \frac{\Gamma(3/4)\Gamma(1/2)}{\Gamma(1/4)}
\frac{N}{T}.
\nonumber
\ee
This dependence of the slope on the temperature $T$ leads to the critical exponent of our plateau transition. Numerically, one can go beyond this approximation and obtain the global behavior of $\sigma^\text{spin}_{xy}$
as a function of $M$. There will be a first-order phase transition which corresponds to the
jump from the black hole embedding to a flat Minkowski embedding.

\section{Field theory of the particle-hole symmetric insulator in $1+1$ dimensions}\label{sec:FT_PHSI}

\subsection{Topological band theory and symmetry classes}
\label{classes}

\begin{table}
\begin{tabular}{c||c|c||ccc}
class & classification & minimal fermion content & $T$ &$C$ &$S$ \\
\hline
\hline
AIII &$\mathbb{Z}$ & 2 Majorana & $0$&$0$&$1$ \\
BDI &$\mathbb{Z}$& 1 Majorana  & $+$&$+$&$1$ \\
CII &$2 \mathbb{Z}$&4 Majorana  &$-$&$-$&$1$\\
\hline
D &$\mathbb{Z}_2$& 1 Majorana  & $0$&$+$&$0$\\
DIII & $\mathbb{Z}_2$& 2 Majorana & $-$&$+$&$1$
 \end{tabular}
\caption{Classification of free fermion systems in $1+1$ dimensions, from Ref.~\onlinecite{schnyder2008,Kitaev2009,schnyder2009,Ryu2009,Ryu:2010hc,Ryu:2010fe} (see main text for explanation).}
\label{TableII}
\end{table}
We now turn our attention towards topological insulators in $1+1$ dimensions. The relevant topological classification for free fermion systems is summarized in Table~\ref{TableII}. The first column gives the name of the symmetry class according to the Altland-Zirnbauer classification~\cite{zirnbauer1992,altland1997}. The second column indicates whether the associated topological invariant is integer-valued ($\mathbb{Z}$), an even integer ($2\mathbb{Z}$), or $\mathbb{Z}_2$-valued. The third column gives the minimal
fermionic matter content~\cite{Ryu:2010hc,Ryu:2010fe} needed to realize the corresponding topological phase. The last three columns indicate whether in this symmetry class a certain discrete symmetry is imposed ($0$ denoting that the symmetry is not present) and if so, whether the matrix of this symmetry action is symmetric ($+$) or antisymmetric ($-$). The three $\mathbb{Z}_2$ symmetries $T$, $C$ and $S$ denote the time-reversal symmetry, the charge conjugation (particle-hole) symmetry, and the sublattice symmetry (or `chiral' symmetry --- not to be confused with the chiral symmetry of massless fermions in even-dimensional spacetimes). In the presence of $C$ and $T$, $S$ is given by their product $S=CT$.
The top three entries (AIII, BDI and CII) are described by an integer topological invariant, while the latter two
(D and DIII) are described by a $\mathbb{Z}_2$ invariant. Classes D and DIII will be the focus of our study. In both cases, the discrete charge conjugation symmetry $C$ ensures that fermions can only become massive in pairs.

To realize a particle-hole symmetric insulator we start from the Dirac fermion action in Minkowski spacetime,
\begin{align}
\label{lag}
\mathcal{L}=\overline{\psi}i\gamma^\mu\partial_\mu\psi-m\overline{\psi}\psi
-im_5\overline{\psi}\gamma^5\psi
\end{align}
with $m$ the normal mass and $m_5$ the axial mass. We have $\{\gamma^\mu,\gamma^\nu\}=2\eta^{\mu\nu}$ and $\eta^{\mu\nu}=\diag(1,-1)$. We choose the Dirac matrices to be all real, $\gamma^0\equiv\sigma_x$, $\gamma^1\equiv-i\sigma_y$, $\gamma^5\equiv\gamma^0\gamma^1=\sigma_z$. The Hamiltonian
\begin{align}
H=-i\gamma^0\gamma^1\partial_1+m\gamma^0+im_5\gamma^0\gamma^5
\nonumber
\end{align}
is Hermitian because $\{\gamma^\mu,\gamma^5\}=0$. The Hamiltonian in momentum space is
\begin{align}\label{Hk}
H(k)=\gamma^0\gamma^1k+m\gamma^0+im_5\gamma^0\gamma^5.
\end{align}
According to the classification of TI~\cite{schnyder2008,Kitaev2009,schnyder2009,Ryu2009}, the $T$ and $C$ symmetries act on $H(k)$ as
\begin{align}
H(k)&=TH^*(-k)T^{-1}, TT^\dag=1, T^T=\eta_TT,\nonumber\\
H(k)&=-CH^*(-k)C^{-1}, CC^\dag=1, C^T=\eta_CC,\nonumber
\end{align}
where $\eta_T=\pm 1$ and $\eta_C=\pm 1$ are given by columns $T$ and $C$ of Table~\ref{TableII}, respectively. To which of the five symmetry classes of Table~\ref{TableII} does the Hamiltonian Eq.~(\ref{Hk}) belong?

\subsubsection{Class D}\label{sec:D}

We first explore the possibility that Eq.~(\ref{Hk}) corresponds to a $\mathbb{Z}_2$ class D TI in $1+1$ dimensions.
In this case, there is no $T$ symmetry but we have a $C$ symmetry with $\eta_C=+1$, hence $C^T=C$ and we need a symmetric, unitary matrix representation of $C$. Since $C=\gamma^0$ does not leave the kinetic term invariant, there are only two choices, the unit matrix $\mathbb{I}$ or $\sigma_z$. Since
\begin{align}
-H^*(-k)=\gamma^0\gamma^1k-m\gamma^0+im_5\gamma^0\gamma^5,
\nonumber
\end{align}
we can choose $C=\mathbb{I}$ if $m=0$, or $C=\gamma^5$ if $m_5=0$. Hence we see that $C^2=+1$, and Eq.~(\ref{Hk}) describes class D.

\subsubsection{Class DIII}\label{sec:DIII}

In this case, in addition to the $C$ symmetry with $\eta_C=+1$ we need a $T$ symmetry with $\eta_T=-1$. This means that $T^T=-T$, i.e. $T$ must be antisymmetric. The only possible choice is $T\propto\sigma_y$, and we can take $T=\gamma^1$ without loss of generality, with $T^2=-1$. We have
\begin{align}
H^*(-k)=-\gamma^0\gamma^1k+m\gamma^0-im_5\gamma^0\gamma^5,
\nonumber
\end{align}
but
\begin{align}
TH^*(-k)T^{-1}=\gamma^0\gamma^1k-m\gamma^0-im_5\gamma^0\gamma^5,
\nonumber
\end{align}
hence the system is $T$-invariant only if $m=0$ {\it and} $m_5=0$.

Physically, this can be understood as the statement that a mass term on the $(1+1)$-dimensional edge of a QSH system necessarily breaks $T$~\cite{Qi2008,qi2008a}. These two masses (normal and axial) can be rotated into one another, generating a charge current~\cite{Qi2008,qi2008a}. In this context, $T^2=-1$ comes from the fact that the right-mover $\psi_R$ and the left-mover $\psi_L$ are time-reversed partners with opposite $z$ component of spin, i.e. $\psi_R\equiv\psi_{R\uparrow}$ and $\psi_L\equiv\psi_{L\downarrow}$. In order to get a gapped system in class DIII without breaking $T$ (if $T^2=-1$), one needs a $4\times 4$ Hamiltonian matrix, e.g. two $2\times 2$ blocks with $T$-breaking masses of different sign for the two blocks. In the continuum description this can simply be implemented by taking {\it two} copies of the simple Dirac fermion described by the action in Eq.~(\ref{lag}) above.

\subsubsection{Insulator versus superconductor realizations}\label{sec:remarks}

Note that the special case of the Hamiltonian (\ref{Hk}) {\it does have} a $T$ symmetry, but with $\eta_T=+1$. Indeed, if we require $T^T=T$, we can in principle take $\mathbb{I}$, $\sigma_z$ or $\sigma_x$. The first two do not leave the kinetic term invariant and are rather $C$ symmetries, as seen in Sec.~\ref{sec:D}. It is not hard to show that $T=\gamma^0$ is a symmetry of Eq.~(\ref{Hk}) with $T^2=+1$. Thus, the statement is that this $T$ symmetry can be present or not, but it does not affect the topological classification into class D (and hence does not appear in Table~\ref{TableII}).

An another way to see that class D is the proper symmetry class for the particle-hole symmetric insulator with action (\ref{lag}) is to perform the dimensional reduction from the $\mathbb{Z}$ class D in $2+1$ dimensions to the $\mathbb{Z}_2$ class D in $1+1$ dimensions~\cite{Ryu2009}. The $\mathbb{Z}$ class D in $2+1$ dimensions corresponds to the $p_x+ip_y$ (or $p_x-ip_y$) chiral superconductor, the Bogoliubov-de Gennes Hamiltonian of which is simply the massive Dirac Hamiltonian in $2+1$ dimensions~\cite{read2000,schnyder2008}. As is well-known, the massive Dirac Hamiltonian in $2+1$ dimensions supports a quantum Hall effect, i.e. the parity anomaly~\cite{Redlich:1983dv,Redlich:1983kn,Semenoff:1984dq,Haldane:1988zz}. Therefore, the class D dimensional reduction from $2+1$ dimensions to $1+1$ dimensions~\cite{Ryu2009} corresponds to the dimensional reduction from the $(2+1)$-dimensional quantum Hall insulator to the $(1+1)$-dimensional $C$-symmetric insulator described in Ref.~\onlinecite{Qi2008}. The idea is well expressed in Ref.~\onlinecite{lee2007}. To each free Majorana fermion model, there is a free complex (Dirac) fermion model with an identical excitation spectrum (modulo the particle-hole redundancy of the spectrum in the Majorana case). The symmetry class D can be realized either by a topological superconductor with a single Majorana fermion~\cite{kitaev2001}, or by a $C$-symmetric TI with a single Dirac fermion. By the same token, the symmetry class DIII can be realized either by a $T$-invariant topological superconductor with two Majorana fermions, or by a $C$- and $T$-(and hence also $S=CT$) invariant TI with two Dirac fermions.

As mentioned in Sec.~\ref{sec:DIII}, to define class DIII we need a $T$-symmetry with $T^T=-T$. For an insulator realization of this class, as explained in Sec.~\ref{sec:D} we need two Dirac fermions which we will denote $\psi_1$ and $\psi_2$, just as in the ($2+1$)-dimensional case of a $T$-invariant QSH insulator. The Hamiltonian matrix will be $4\times 4$.  As seen in Sec.~\ref{sec:DIII}, the $T$-symmetry $T=\gamma^1$ flips the sign of both masses (normal and axial). However we require $m_5=0$ if the system is to be $C$-invariant with $C=\gamma^5$ (Sec.~\ref{sec:D}). Therefore we can choose $m$ for $\psi_1$ and $-m$ for $\psi_2$, i.e.
\begin{align}
H(k)=\left(
\begin{array}{cc}
\gamma^0\gamma^1k+m\gamma^0 & 0 \\
0 & \gamma^0\gamma^1k-m\gamma^0
\end{array}
\right),\nonumber
\end{align}
with the symmetries $C=\mathbb{I}\otimes\gamma^5$ and $T=\gamma^0\otimes\gamma^1$. The $\gamma^0$ in $T$ simply interchanges the fermions with $m$ and $-m$, to preserve the $T$-invariance. It is also easy to check that $T^T=-T$ and $T^2=-1$. The story is thus very much like in the QSH effect, where fermions of opposite spin have opposite signs of the mass. We can write the second-quantized Hamiltonian as
\begin{align}
\mathcal{H}(k)=\Psi^\dag_k\left(\gamma^0\gamma^1k+m\gamma^5\otimes\gamma^0\right)\Psi_k,
\nonumber
\end{align}
where $\Psi_k^\dag=\left(\begin{array}{cc}\psi^\dag_{k1} & \psi^\dag_{k2}\end{array}\right)$.

\subsection{Topological field theory and axial anomaly}

\subsubsection{Class D}

Consider a single flavor of massive Dirac fermions in $1+1$ dimensions as in Eq.~(\ref{lag}) above with an axial mass angle
$\theta$, and couple it to the $U(1)$ electromagnetic gauge potential $A_\mu$,
\begin{align}\label{L1+1Minkowski}
\mathcal{L}=\overline{\psi}i\gamma^\mu(\partial_\mu+ieA_\mu)\psi
-m\cos\theta\overline{\psi}\psi-im\sin\theta\overline{\psi}
\gamma^5\psi,
\end{align}
where we have redefined the normal and axial masses $m$ and $m_5$ in Eq.~(\ref{lag}) as $m\cos\theta$ and $m\sin\theta$, respectively. As we have seen in Sec.~\ref{sec:D}, the Lagrangian $\mathcal{L}$ is invariant under charge
conjugation only if the axial mass vanishes, that is if $\theta=0$ or $\theta=\pi$ (modulo a
periodicity of $2\pi$ in $\theta$). We will now show that the
effective action for the electromagnetic field
$S_\mathrm{eff}[A_\mu,\theta]$ obtained by integrating out the
fermions in Eq.~(\ref{L1+1Minkowski}) contains a topological term
of the form $S_\mathrm{eff}[A_\mu,\theta]=\frac{\theta
e}{4\pi}\epsilon^{\mu\nu}F_{\mu\nu}$, which
establishes Eq.~(\ref{L1+1Minkowski}) as the continuum field theory of
particle-hole symmetric insulators in $1+1$ dimensions, with
$\theta=\pi$ corresponding to the $\mathbb{Z}_2$ insulator and
$\theta=0$ to the trivial insulator.

The derivation of the topological term is almost identical to the one in the ($3+1$)-dimensional case and also is governed by an anomaly. It is most convenient in 2-dimensional Euclidean signature.
The action is quadratic in fermions, so the Euclidean
functional integral can be performed
\begin{align}
Z=\int\mathscr{D}\psi\mathscr{D}\overline\psi\mathscr{D}A_\mu
\,e^{-\int d^2x\,\mathcal{L}}=\int\mathscr{D}A_\mu\,
e^{-S_\mathrm{eff}[A_\mu,\theta]},\nonumber
\end{align}
where the effective action $S_\mathrm{eff}[A_\mu,\theta]$ for the
electromagnetic field $A_\mu$ is defined as
\begin{align}
e^{-S_\mathrm{eff}[A_\mu,\theta]}\equiv
\int\mathscr{D}\psi\mathscr{D}\overline\psi
\,e^{-\int d^2x\,\mathcal{L}}.\nonumber
\end{align}
In order to calculate the effective action $S_\mathrm{eff}$, one first performs a chiral rotation to eliminate the mass angle
$m\overline\psi e^{i\theta\gamma_5}\psi\rightarrow
m\overline\psi\psi$. In so doing, one observes that due to the
nontrivial transformation of the functional integration measure
$\mathscr{D}\psi\mathscr{D}\overline\psi$ under the chiral
rotation (which manifests itself as an axial anomaly in the
massless case~\cite{Fujikawa:1979ay,Fujikawa:1980eg}), the action acquires a
topological term~\cite{Kikuchi:1992rz} $S\rightarrow S+\frac{i\theta
e}{4\pi}\epsilon_{\mu\nu}F_{\mu\nu}$. The fermions now have a real positive mass and can be
safely integrated out, yielding a contribution to the effective
action for the electromagnetic field which has no $\theta$
dependence. The only difference in $1+1$ dimensions is that the group theory coefficient $C$ determining the anomaly for a general matter content is the sum of the electric charge over all fields (times their axial charge, which we took to be 1 in the above discussion), as opposed to the sum of the \emph{square} of the electric charge in $3+1$ dimensions. This is due to the fact that in $3+1$ dimensions the anomaly is given by triangle diagrams, whereas in $1+1$ dimensions it comes from bubble diagrams.
So even in the fractional case where we have $N$ partons of charge $1/N$, the effective $\theta$ angle is still $\theta=\pi$ when all partons are in the topologically non-trivial phase. The fractionalized state is still a topologically distinct phase; the $\theta$ angle alone however is insensitive to the difference. In the AdS/CFT normalization where each parton has charge 1 and the electron has charge $N$, we once more predict a $\theta$ angle of $\theta=N$, which is of the same order (in $N$) as in the ($3+1$)-dimensional case and hence comes from a classical calculation in the holographic dual.

One physical consequence of the $\theta$ angle is that at an interface between a topologically non-trivial region and a topologically trivial one (e.g. vacuum) a constant charge will be induced~\cite{Qi2008}. As the field strength $F=dA$ is a total derivative, the $\theta$ term can be integrated to a boundary term localized at the interface across which $\theta$ jumps by $\pi$,
\be\label{e/2charge}
S_\text{interface} = \pm \frac{e}{2} \int dt \, A_t,
\ee
corresponding to an induced electric charge of $\pm \frac{e}{2}$.

\subsubsection{Class DIII}\label{sec:classDIII}

As we have seen in Sec.~\ref{sec:D}, to describe a class DIII particle-hole symmetric insulator in $1+1$ dimensions we have to start with two copies of the Dirac Lagrangian Eq.~(\ref{lag}) for fermions $\psi_1$ and $\psi_2$. A $T$-invariant mass combines a real, positive mass for $\psi_1$ with a real, negative mass of equal magnitude for $\psi_2$ in the topologically trivial case. The topologically non-trivial insulator corresponds to the opposite choice of sign (positive mass for $\psi_2$ and negative mass for $\psi_1$). The Lagrangian is simply the dimensional reduction of the ($2+1$)-dimensional Lagrangian we used for the QSH effect [Eq.~(\ref{freeaction}),(\ref{threedmass})]. Again, $\psi_1$ and $\psi_2$ carry the same charge under the Maxwell $U(1)$ field. As we have doubled the matter content, the free Lagrangian has a $U(2)$ global symmetry which is broken to $U(1)_\text{Maxwell}\times U(1)_R$ after gauging electromagnetism. $U(1)_R$ is a global symmetry under which $\psi_1$ carries charge $+1$ and $\psi_2$ carries charge $-1$. As in the case of $2+1$ dimensions, this is the continuum field theory realization of spin.

The anomaly calculation completely parallels the one of the previous subsection. However, in the present case, in both the topologically trivial and the topologically non-trivial insulator we have exactly one negative mass fermion of Maxwell charge 1. Therefore, in both phases we generate a $\theta$ term with $\theta=\pi$. Correspondingly, there is no jump in $\theta$ when crossing an interface between the two, and hence also no induced electric charge.
Using the same anomaly technique we can also look for the generation of a $\theta F^R$ term, where $F^R$ is the field strength of a background $U(1)_R$ gauge field. As $\psi_1$ has positive charge under the $R$-symmetry while $\psi_2$ has negative charge, we will this time generate $\theta=-\pi$ in the trivial case and $\theta=\pi$ in the non-trivial case. While $\theta$ is only defined up to shifts of $2 \pi$, in the case of an interface the sign difference is nonetheless physical as only the {\it overall} $\theta$ angle can be shifted by $2 \pi$. When adding $2 \pi$ to $\theta$ to shift $\theta=-\pi$ to $\theta=\pi$ in the trivial insulator, the $\theta$ angle in the non-trivial insulator becomes $3 \pi$. Due to the $2 \pi$ jump in $\theta$ across the defect we induce an integer $U(1)_R$ charge. That is, there is an induced unit spin (in units of $\frac{\hbar}{2}$) on the defect. One can arrive at the same conclusion
by performing the Jackiw-Rebbi analysis~\cite{Jackiw:1975fn} and considering a mass kink $m(x)$. Since the masses of the $\psi_1$ and $\psi_2$ fermions have opposite signs, we will have a localized charge $\frac{e}{2}-\frac{e}{2}=0$ and a localized spin $\frac{\hbar}{4}+\frac{\hbar}{4}=\frac{\hbar}{2}$, i.e. a spinon. This is not a true $SU(2)$ spin but rather a $U(1)$ spin, with the spin generator $S_z=\frac{\hbar}{2}\gamma^5\otimes\mathbb{I}$. This is a dimensionally reduced version of the phenomenon of spin-charge separation in the QSH insulator~\cite{qi2008c,ran2008}.

\section{Holographic fractional particle-hole symmetric insulator in $1+1$ dimensions}\label{sec:H_PHSI}

\subsection{Class D}

For the $T$-invariant TI in $3+1$ dimensions and the QSH effect in $2+1$ dimensions, the minimal matter content for a single flavor of topologically non-trivial fermions essentially corresponded to a single ($3+1$)-dimensional Dirac fermion. This is the fermionic content of a hypermultiplet in supersymmetric gauge theories with $8$ supercharges ($\mathcal{N}=2$), and correspondingly it was possible to find a holographic realization of these TI states in terms of an intersecting brane setup with 4 ND directions. The ($1+1$)-dimensional particle-hole symmetric insulator in class D only requires a single ($1+1$)-dimensional Dirac fermion, which is the dimensional reduction of a single ($3+1$)-dimensional Weyl fermion, that is, half of the matter in a $\mathcal{N}=2$ hypermultiplet. Instead, we are looking for the dimensional reduction of a single ($3+1$)-dimensional chiral multiplet of a theory with $4$ supercharges ($\mathcal{N}=1$).

Brane setups realizing the fermionic matter content of a ($3+1$)-dimensional chiral multiplet on their intersection
have 6 ND directions. They are generically non-supersymmetric and hence unstable. Indeed, for the simplest 6ND system along the lines above, a probe D5-brane on AdS$_3 \times S^3$ inside the AdS$_5 \times S^5$ background supported by $N$ D3-branes (this is the near horizon limit of D3-branes along $0123$ and D5-branes along $014567$), the slipping mode that contracts the $S^3$ wrapped by the D5-brane has a mass squared of $-3$. This is below the Breitenlohner-Freedman bound of $-1$ appropriate for AdS$_3$, and hence signals an instability. However, 6ND setups can be stabilized by taking the branes to intersect at an angle and/or turning on appropriate background gauge fields on the brane. They can even be supersymmetric. It is well known how to do this in flat space~\cite{Berkooz:1996km}. Some aspects of the corresponding mechanism in the near horizon geometry have been elucidated in Ref.~\onlinecite{myers2008,Bergman:2010gm}. Fortunately for us, the details of the stabilization mechanism do not matter. The calculation of the boundary term proceeds as in Ref.~\onlinecite{toappear}. As long as we assume that there is some brane configuration which describes massive fundamental flavors, the mass of which changes sign at $x=0$ ($x$ being the spatial direction of the field theory), we can calculate the induced charge from the WZ term without knowing any other details of the brane embedding.

Using the metric on $S^5$ from Eq.~(\ref{ofermetric}), we can use $t$ and the five coordinates on $S^5$ to parametrize the worldvolume of the D5-brane. In order to solve for the embedding of the D5-brane, one conventionally would take $x$ and $r$ as worldvolume coordinates and then solve for the angles $\theta$ and $\psi$ as functions of $x$ and $r$ to characterize the embedding. For the D3-D7 system this has been done explicitly via a numerical calculation~\cite{toappear}. Using directly $\theta$ and $\psi$ as worldvolume parameters, the integral over spatial coordinates just becomes the integral over a part of the $S^5$ and all we need to determine is the range the angles take on the solution. For a massive embedding, the $\theta$ angle covers its full range from $0$ to $\pi$. $\psi$ directly corresponds to the phase of the mass term. It changes from $0$ to $\pi$ as we go from $x=-\infty$ to $x=\infty$ (or vice versa), that is, in the solution it covers {\it half} of its $2 \pi$ range. The relevant WZ term is once more given by the action in Eq.~(\ref{conventions}). Recall that the net 5-form flux over the $S^5$ is given by~\cite{fluxconvention}
\be
\int_{S^5} dC_4 = (2 \pi)^4 (\alpha')^2 \, N.
\nonumber
\ee
Also, we note that the integral to be done only differs from the integral over the full $S^5$ by integrating $\phi$ over $\Delta \phi=\pi$, the range of values realized in the solution, as opposed to its full $2 \pi$ range~\cite{toappear}. In other words, the integral goes over half the sphere. The 5-form flux is proportional to the volume form $d\text{vol}_5$ on the $S^5$, which is independent of $\phi$. Therefore, we just pick up half the flux from the integral over the worldvolume, and we conclude
\begin{eqnarray*}
\nonumber
S_\text{WZ} &=& (2 \pi \alpha') T_5 \int dt d\text{vol}_5 F \wedge C_4 \\
&=&
(2 \pi \alpha') T_5 \left(\int A_t dt\right) \left(\int dC_4\right) \\
&=& \frac{N}{2} \int A_t dt,
\end{eqnarray*}
in perfect agreement with our field theory analysis Eq.~(\ref{e/2charge}). Locally, close to the intersection, the D5-brane looks like the 8ND AdS$_2 \times S^4$ embedding studied recently (e.g. Ref.~\onlinecite{Kachru:2009xf}), and the appearance of the half unit of charge on the interface is essentially the well-known (in the stringy literature) Hanany-Witten effect~\cite{Hanany:1996ie}.

\subsection{Class DIII}

With its doubled matter content, the class DIII particle-hole symmetric insulator in $1+1$ dimensions does correspond to the fermionic content of a $\mathcal{N}=2$ hypermultiplet and can be realized by a supersymmetric 4ND brane setup, an intersecting D3-D3 system with the probe D3-brane wrapping an asymptotically AdS$_3 \times S^1$ geometry. We can once more compute the induced charges on an interface between topologically trivial and non-trivial insulators. As the worldvolume is only 4-dimensional, we clearly cannot have any non-trivial $F \wedge C$ WZ terms this time. This is consistent with our field theory calculation indicating that there is no induced charge on the defect for class DIII. We expect however to have a $N \int dt \, A_t^R$ term localized on the interface and corresponding to the localized spin. For the massive embedding, the D3-brane worldvolume sweeps out an $S^2$ inside the $S^5$. For a D3-brane interpolating between positive mass at large $x$ and negative mass at small $x$, we in addition sweep out half of the analog of the $\phi$ direction. As a result, this embedding will cover half an $S^3$ inside the $S^5$. As above, instead of using $r$ and $x$ as worldvolume coordinates, we directly use the three angles on this $S^3$ together with $t$ to parametrize our worldvolume. We can once more write the internal $S^5$ metric as in Eq.~(\ref{ofermetric}), as well as identify the $R$-gauge field via Eq.~(\ref{oferrfield}) with the normalization given by Eq.~(\ref{oferconvention}). We finally obtain
\be
S_\text{WZ}= \tilde{\eta} T_{3} \pi^2 \, \int dt \, A_t^R,
\nonumber
\ee
where the factor of $\pi^2$, as in the case of the ($2+1$)-dimensional Chern-Simons term (\ref{CSterm}) for the quantum $R$-Hall effect, is the volume of the 3-hemisphere. Using Eq.~(\ref{oferconvention}) together with
\be
T_3 = \frac{1}{(2\pi)^3 (\alpha')^2},
\nonumber
\ee
one finds
\be
S_\text{WZ}= N \int dt A_t^R,
\nonumber
\ee
as expected from the field theory results of Sec.~\ref{sec:classDIII}.

\section{Conclusion}

In this work, we constructed low-energy effective field theories for fractional versions of the $T$-invariant TI in $2+1$ dimensions, i.e. the QSH insulator, and the $C$-invariant TI in $1+1$ dimensions, in terms of massive Dirac fermions. Integrating out the massive fermions, we derived topological field theories describing the electromagnetic and spin response of these states. We then constructed holographic realizations of these states in terms of D-brane systems, where WZ terms in the brane actions were shown to reproduce the results of the topological field theory description.

In $2+1$ dimensions, assuming that the $z$ component of spin was conserved in the condensed matter system, the associated $U(1)_R$ symmetry allowed us to derive a topological mixed Chern-Simons term via the parity anomaly, which described the QSH effect as a quantum Hall effect of $R$-charge. The fractional QSH state was obtained by breaking up the electron of charge $1$ into $N$ partons of charge $1/N$, which resulted in a fractional Chern-Simons level $1/N$ corresponding to a fractional QSH effect. We then obtained a microscopic holographic realization of this fractional QSH state in terms of probe D5-branes in an AdS$_5\times S^5$ background. The quantization of the spin Hall effect followed directly from the topological nature of the WZ terms in the brane action and was seen to be independent of the details of the brane embedding. In $1+1$ dimensions, the effective field theory of the class D and class DIII TI was found to be a theory of one and two flavors of Dirac fermions, respectively. For class D, the Maxwell $U(1)$ symmetry allowed us to derive a topological $\theta$-term via the chiral anomaly, which described the existence of a charge $\pm e/2$ fermion zero mode on a domain wall of $\theta$ with $\theta(x=+\infty)-\theta(x=-\infty)=\pm \pi$. For class DIII, the topological term was a $U(1)_R$ $\theta$-term describing an electrically neutral localized spin (i.e. a spinon). In $1+1$ dimensions, the parton construction did not give a fractional $\theta$, but the fractionalization of the electron into $N$ deconfined partons did nonetheless distinguish the fractional TI from the noninteracting TI. We obtained a microscopic holographic realization of the fractional class D TI in terms of a stabilized 6ND brane setup, where the details of stabilization mechanism were not needed to derive the topological WZ term. For the fractional class DIII TI, the holographic realization was in terms of a supersymmetric 4ND setup.

\section*{Acknowledgements}

We would like to thank O. Bergman, D. L. Harlow, X. L. Qi, S. Ryu and S. Yaida for helpful discussions. AK is supported in part by DOE grant DE-FG02-96ER40956. JM is supported by a Stanford Graduate Fellowship. TT is supported by World Premier International Research Center Initiative (WPI Initiative), MEXT, Japan. The work of TT is also supported in part by JSPS Grant-in-Aid for Scientific Research No.~20740132, and by JSPS Grant-in-Aid for Creative Scientific Research No.~19GS0219.

\bibliography{holofti}

\begin{thebibliography}{72}
\bibitem[{Rsy({\natexlab{a}})}]{Rsymmetry}
\bibinfo{note}{In a condensed matter context where Lorentz invariance is not
  required, it is possible to add terms which break the $R$-symmetry without
  violating $T$-invariance. Physically, these terms originate from the breaking
  of inversion symmetry due to the crystal structure or background electric
  fields. Here we only consider the $R$-symmetric case, which in the condensed
  matter language corresponds to the conservation of the $z$ component of
  spin.}
\bibitem[{nor()}]{normalization}
\bibinfo{note}{In Ref.~\onlinecite{Aharony:1999rz} the authors find the
  corresponding normalization constant to be $\eta=1/2$, but this was obtained
  after the form field was rescaled by a factor $16 \pi \alpha'^2 (2
  \tilde{N})$ where $2 \tilde{N}=N$ was introduced as they were studying a
  $\mathbb{Z}_2$ orientifold geometry. We included this full normalization
  factor in $\tilde{\eta}$.}
\bibitem[{flu()}]{fluxconvention}
\bibinfo{note}{Note that compared to Ref.~\onlinecite{toappear}, in this work
  we use a RR 4-form $C_4$ that is rescaled by a factor of $g_s$.}
\end{thebibliography}

\end{document}